\documentclass[superscriptaddress,12pt]{revtex4}
\usepackage{amsfonts}
\usepackage{graphicx, amsmath}
\usepackage{subfigure}
\begin{document}

\title[Short Title]{Simple implementation of quantum delayed-choice experiment using conventional linear optical elements}
\author{Qi Guo}
\affiliation{Department of Physics, Harbin Institute of Technology,
Harbin, Heilongjiang 150001, People's Republic of China}
\author{Liu-Yong Cheng}
\affiliation{Department of Physics, Harbin Institute of Technology,
Harbin, Heilongjiang 150001, People's Republic of China}
\author{Hong-Fu Wang}
\affiliation{Department of Physics, College of Science, Yanbian
University, Yanji, Jilin 133002, People's Republic of China}
\author{Shou Zhang\footnote{E-mail: szhang@ybu.edu.cn}}
\affiliation{Department of Physics, Harbin Institute of Technology,
Harbin, Heilongjiang 150001, People's Republic of China}
\affiliation{Department of Physics, College of Science, Yanbian
University, Yanji, Jilin 133002, People's Republic of China}

\begin{abstract}
We propose a simple implementation scheme of quantum
delayed-choice experiment in linear optical system
without initial entanglement resource. By choosing
different detecting devices, one can selectively observe the
photon's different behaviors after the photon has been passed the
Mach-Zehnder interferometer. The scheme shows that the photon's
wave behavior and particle behavior can be observed with a single
experimental setup by postselection, that is, the photon can show
the superposition behavior of wave and particle. Especially, we
compare the wave-particle superposition behavior and the
wave-particle mixture behavior in detail, and find the
quantum interference effect between wave and particle behavior, which
may be helpful to reveal the nature of photon essentially.
\end{abstract}
\maketitle

\section{Introduction}

For centuries, scientists have always debated whether the light is
particle or wave. Now it is well accepted that light is both
particle and wave, that is, wave-particle duality. However, Bohr's
principle of complementarity\cite{1} shows that no experiment can
measure both the wave and the particle behaviors simultaneously,
which plays an important role in quantum physics. Mach-Zehnder
interferometer (MZI) is an effective tool for verifying
wave-particle duality, because when light passes through MZI, one
can obtain interference pattern in both of the outputs of the MZI,
and the light behaves as wave. While, if the second beam splitter
(BS) (or say output BS) of the MZI is removed, we'll know the
explicit which-path information of photons and can't obtain
interference pattern in the outputs, so the light behaves as
particle. However, another objection is that maybe photons somehow
know whether the second BS is removed or not in advance, so they
can adjust themselves to the corresponding device. In order to
test this conjecture, Wheeler proposed a famous thought
experiment, i.e. delayed-choice experiment\cite{2,3}. In this
experiment, observers can decide to observe the wave or particle
behavior of a photon after it passed the first BS of a MZI by
inserting or removing the second BS, so that the photon can't know
what measuring devices lie ahead before it is emitted. This
gedanken experiment has attracted great attention\cite{4,5,6,7}.
With the development of experimental technology, some interesting
thought experiments can be realized in current laboratory. Jacques
\emph{et al.} performed Wheeler's delayed-choice experiment with
optical interferometer\cite{8} and tested quantum complementarity
with interfering single photons\cite{9}.

In 2011, Ionicioiu \emph{et al.} presented a quantum version
delayed-choice experiment by replacing the second BS with a
quantum-controlled BS (q-BS)\cite{10}, which could delay the
choice of observers until the photon left the MZI. This is a
significant progress from classical control to quantum control, so
it has attracted the attention of many researchers and some
experimental realizations have been completed in different
physical systems\cite{11,12,13,14,15,16}. These works amply
demonstrated the validity of the quantum delayed-choice
experiment and provided effective methods for testing the particle
and wave behaviors of photons. In addition, quantum delayed-choice
experiment opens the probability for generating photon's wave-particle
superposition state, in other words, quantum delayed-choice experiment
allows wave and particle behavior to be observed simultaneously in a
single experimental setup. Inspired by Ref.~[15], we here present a simple
implementation scheme of quantum delayed-choice experiment by using the commonest linear optical
elements without initial entangled photon pairs. By choosing corresponding detecting device after the photon left
the MZI, one can selectively observe photon's wave behavior, particle behavior,
wave-particle mixture behavior and wave-particle superposition behavior. they only observed the
statistical mixture of wave and particle behavior. More importantly, we show the genuine superposition
behavior of wave and particle state, and compare it with the wave-particle mixture behavior revealed in the existing theoretical and experimental works.

\section{Quantum version of delayed-choice experiment}

Now we briefly review the quantum delayed-choice experiment. The
main difference between quantum and classical delayed-choice
experiment is that the second BS of the MZI is replaced by a
quantum-controlled BS in the former, which means the presence or
absence of the second BS is controlled by a quantum superposition
state rather than a experimenter or a classical random number
generator\cite{10}. For simplicity, we explain the quantum
delayed-choice experiment with Fig.~1(a). The input photon is in
the superposition state of horizontal and vertical polarization,
i.e. $|\varphi\rangle=\sin\alpha|H\rangle+\cos\alpha|V\rangle$,
$|H\rangle$ and $|V\rangle$ denote the horizontal and vertical
polarization state of the photon, respectively. The two paths of
the MZI express as $|0\rangle$ and $|1\rangle$. Here, q-BS is a
polarization-dependent BS, which can completely transmit
vertically polarized photons and 50/50 split horizontally
polarized photons (the realization of q-BS will be introduce in
the next section). Therefore, a vertically polarized photon passes
one path and exhibits particle behavior, but an horizontally
polarized photon passes both of two paths and exhibits wave
behavior. If we consider the polarization and path degree of
freedom as control and target qubit, respectively, the equivalent
quantum circuit diagram of Fig.~1(a) can be represented as
Fig.~1(b).  $\theta$ is a single-qubit gate
$U(\theta)=\mathrm{diag}(1, e^{i\theta})$. The detector $D_{c}$
can distinguish the control qubits $|0\rangle$ and $|1\rangle$
(here $|0\rangle\equiv|V\rangle$ and $|1\rangle\equiv|H\rangle$),
and $D_{t}$ represents the device of detecting the particle or
wave behavior of the photon, which means the devices contains not
only the photon detectors but also the analysis procedure of the
coincidence count probabilities after the detecting.

\begin{figure}
\scalebox{0.9}{\includegraphics{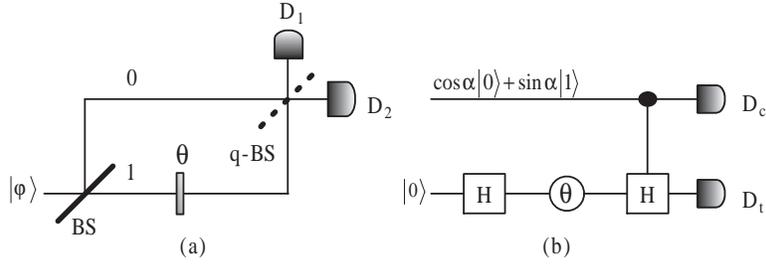}}\caption{\label{f1} (a)
Schematic of the quantum delayed-choice experiment. BS is a beam
splitter. $\theta$ is a the relative phase difference between the
two paths. q-BS is quantum BS, which means the second BS is
inserted for horizontally polarized photons and removed for
vertically polarized photons. $\mathrm{D_{1}}$ and
$\mathrm{D_{2}}$ are photon detectors. (b)The equivalent quantum
circuit diagram of (a). The control qubit is served as by the
polarization degree of freedom of the photon in (a). H denotes
Hadamard conversion. $\mathrm{D_{c}}$ can distinguish the states
of the control qubit $|0\rangle$ and $|1\rangle$, and
$\mathrm{D_{t}}$ is the device of detecting the particle or wave
behavior of the photon.}
\end{figure}

In original Wheeler's delayed-choice experiment, the decision of
inserting the second BS or not must be made after the photon
passes the first BS and before it reaches the second BS. However,
this temporal arrangement is not necessary anymore in the quantum
delayed-choice experiment. For the initial state
$|\phi\rangle=(\cos\alpha|0\rangle+\sin\alpha|1\rangle)|0\rangle$,
the control qubit of Fig.~1(b) will be entangled with the behavior
of the input photon after passing through the quantum network,
which can be expressed as
\begin{eqnarray}\label{e1}
|\phi'\rangle=\cos\alpha|0\rangle|\mathrm{particle}\rangle+\sin\alpha|1\rangle|\mathrm{wave}\rangle,
\end{eqnarray}
where
$|\mathrm{particle}\rangle=\frac{1}{\sqrt{2}}(|0\rangle+e^{i\theta}|1\rangle)$
and
$|\mathrm{wave}\rangle=e^{i\theta/2}(\cos\frac{\theta}{2}|0\rangle-i\sin\frac{\theta}{2}|1\rangle)$
represent the particle and wave behavior\cite{10}, respectively.
Therefore, the detection results of $D_{c}$ and $D_{t}$
necessarily relate to each other. If we know one of the two
detection devices' results, the state of Eq.~(1) will collapse and
the result of the other detecting device will be immediately
determined even without working. Hence, the choice of observing a photon's
wave or particle behavior can be delayed to the future light cone
of the event that the photon leaves the MZI, and can't be made in
the past light cone of the detection of the photon.

\section{Implementation scheme and wave-particle superposition}

In Fig.~2, we show how to simply realize quantum delayed-choice
experiment with ordinary optical elements. Compared with
the existing works, our scheme is simpler and doesn't need initial
entanglement resource. Moreover, we will research the genuine
superposition behavior of wave and particle state, and clearly
compare it with the classical mixture of wave and particle
behaviors.

\begin{figure}
\scalebox{0.9}{\includegraphics{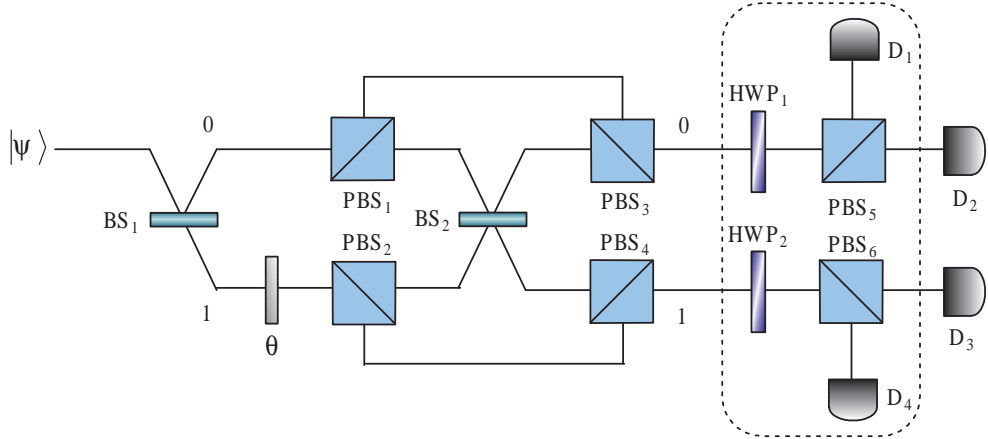}}\caption{\label{f2} The
implementation setup for quantum delayed-choice experiment.
$|\psi\rangle=\sin\alpha|H\rangle+\cos\alpha|V\rangle$. BS: beam
splitter. 0 and 1 indicate the two paths of the MZI. PBS:
polarization beam splitter. $\theta$: adjustable phase shifter.
HWP: half wave plate oriented at $22.5^{\circ}$. $\mathrm{D_{i}}$:
photon detector. If the device in rounded rectangle is removed,
the detectors $\mathrm{D_{2}}$ and $\mathrm{D_{3}}$ will detect
the classical mixture of wave and particle behavior. If the device
in rounded rectangle is inserted, $\mathrm{D_{1}}$ and
$\mathrm{D_{4}}$ ($\mathrm{D_{2}}$ and $\mathrm{D_{3}}$) will
detect superposition behavior of wave and particle state, but when
we remove the HWPs, $\mathrm{D_{1}}$ and $\mathrm{D_{4}}$
($\mathrm{D_{2}}$ and $\mathrm{D_{3}}$) will detect particle
(wave) behavior respectively. }
\end{figure}

The initial photon is in the polarization superposition state
$|\psi\rangle=\sin\alpha|H\rangle+\cos\alpha|V\rangle$. The two
paths also express as $|0\rangle$ and $|1\rangle$, respectively.
Let the photon enter the setup from path 0. After the photon
passes though the $\mathrm{PBS_{3}}$ and $\mathrm{PBS_{4}}$, the
final state can be expressed with both the path degree of freedom
and the polarization degree of freedom
\begin{eqnarray}\label{e2}
|\psi'\rangle=\cos\alpha|\mathrm{particle}\rangle|V\rangle+\sin\alpha|\mathrm{wave}\rangle|H\rangle,
\end{eqnarray}
where $|\mathrm{particle}\rangle$ and $|\mathrm{wave}\rangle$ are
the same as Eq.~(1). Equation~(2) actually is a single photon
path-polarization hyperentangled state. Then we can detect
different behaviors of the photon by choosing different detection
devices. If we remove the two HWPs in Fig.~2, obviously,
$\mathrm{D_{1}}$ and $\mathrm{D_{4}}$ detect the particle
behavior, whose probabilities of detecting photon are
$I_{1}=I_{4}=\frac{1}{2}$, resulting in the visibility of the
interference pattern is
$\mathcal{V}_{1}=\mathcal{V}_{4}=(I_{max}-I_{min})/(I_{max}+I_{min})=0$.
And $\mathrm{D_{2}}$ and $\mathrm{D_{3}}$ detect the wave
behavior, so the detecting probabilities of $\mathrm{D_{2}}$ and
$\mathrm{D_{3}}$ are $I_{2}=\cos^{2}\frac{\theta}{2}$ and
$I_{3}=\sin^{2}\frac{\theta}{2}$, respectively, resulting in the
visibility $\mathcal{V}_{2}=\mathcal{V}_{3}=1$. If we remove the
device in rounded rectangle of Fig.~2 and directly detect the
photon in the two paths with $\mathrm{D_{2}}$ and
$\mathrm{D_{3}}$, we can obtain the mixture state of particle and
wave
\begin{eqnarray}\label{e3}
\rho=\mathrm{Tr}_{pol}|\psi'\rangle\langle\psi'|=\cos^{2}\alpha|\mathrm{particle}\rangle\langle\mathrm{particle}|+\sin^{2}\alpha|\mathrm{wave}\rangle\langle\mathrm{wave}|,
\end{eqnarray}
which is the reduced density matrix tracing out the polarization
qubit for Eq.~(2). Take $\mathrm{D_{2}}$ for example, the
probability that $\mathrm{D_{2}}$ detects photon (or interference
pattern) for the mixture state is
\begin{eqnarray}\label{e4}
I_{0}(\theta,\alpha)=\mathrm{Tr}[\rho|0\rangle\langle0|]
=\cos^{2}\frac{\theta}{2}\sin^{2}\alpha+\frac{1}{2}\cos^{2}\alpha.
\end{eqnarray}
We plot the graph of the above probability distribution function
$I_{0}(\theta,\alpha)$ as shown in Fig.~3(a). It's easy to see
that Fig.~3(a) is the same as the previous works in
Refs.~[10,14,15], which indicates the wave-particle mixture state has been revealed and experimental results fitted well with theoretical predictions.

\begin{figure}
\renewcommand\figurename{\small Fig.}
 \centering \vspace*{8pt} \setlength{\baselineskip}{10pt}
 \subfigure[]{
 \includegraphics[scale = 0.5]{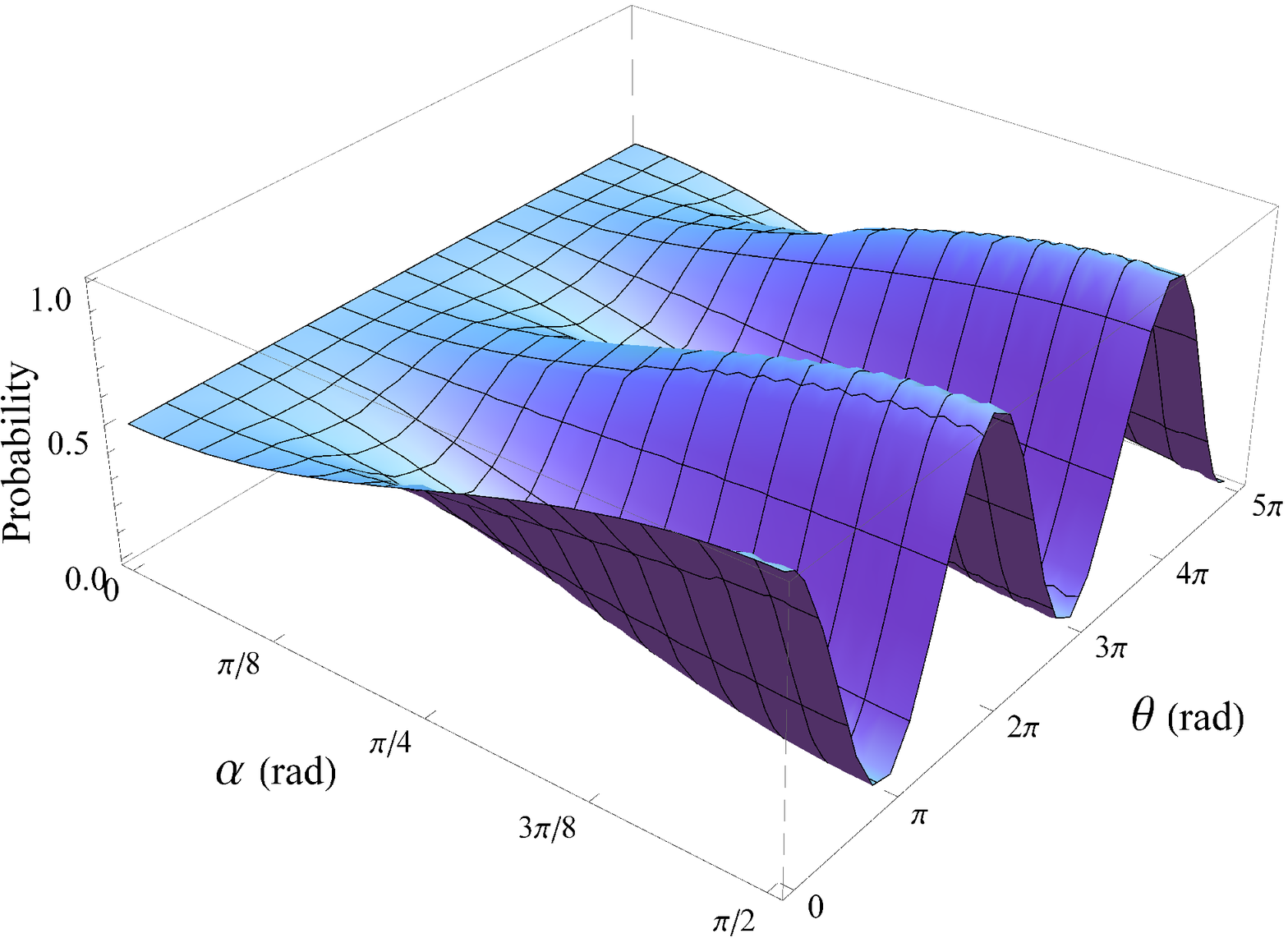}}
 \subfigure[]{
 \includegraphics[scale = 0.5]{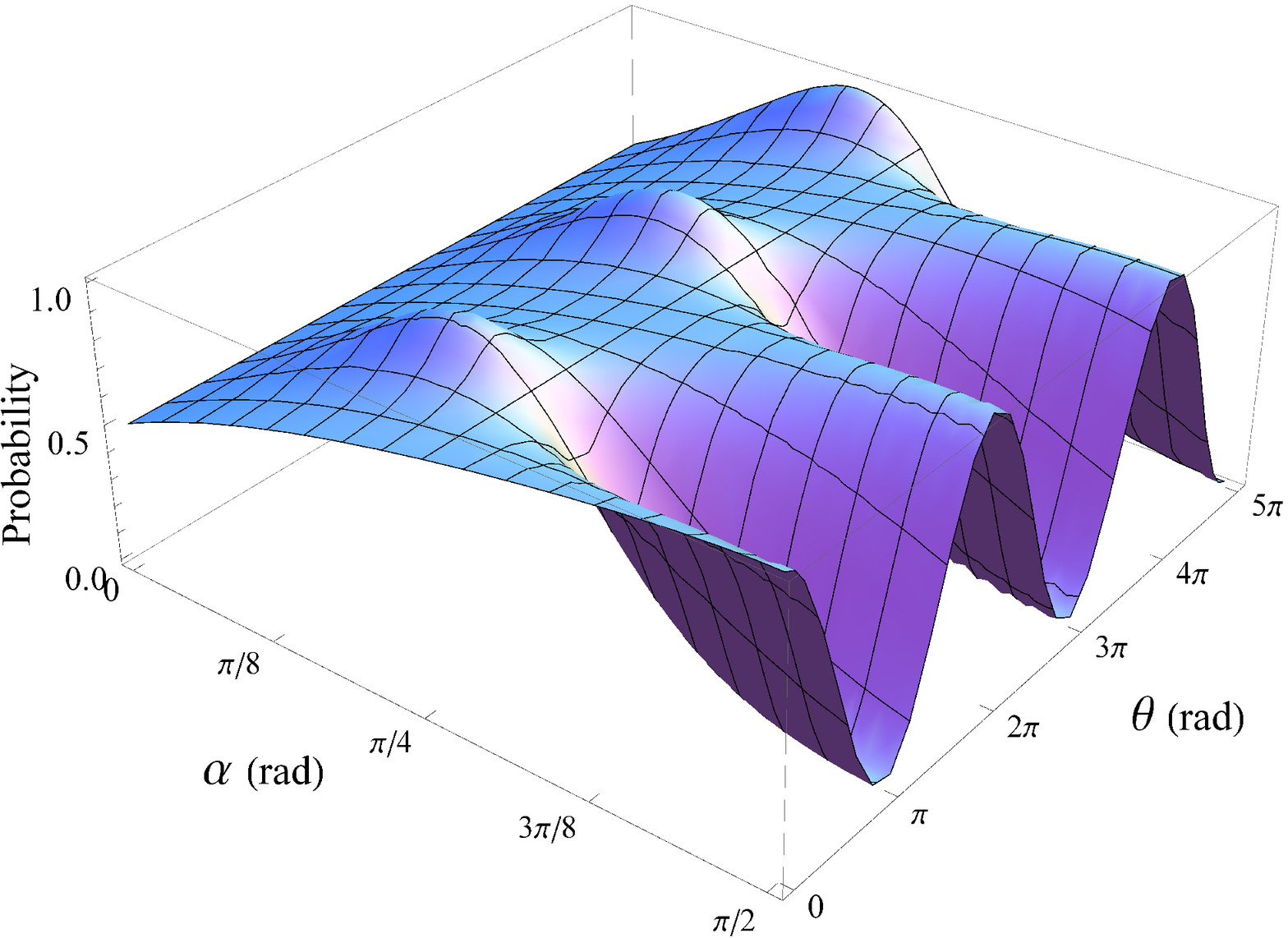}}
 \caption{\label{f3} Plots of the probability distributions of the photon in path 0 (detector $\mathrm{D_{2}}$) of the Fig.~2.
For clarity, we choose $0<\theta<5\pi$. (a) shows the mixture
behavior of wave and particle, as defined by Eq.~(4),
$I_{0}(\theta,\alpha)$. (a) is the same as the Fig.~3(A) and (B)
in Ref.~\cite{14}, Fig.~2 in Ref.~\cite{10}, and Fig.~4(A) and (C)
in Ref.~\cite{15}. (b) is the genuine wave-particle superposition behavior, as defined by
Eq.~(7), $I'_{0}(\theta,\alpha)$. Because of the coherent quantum
superposition of wave and particle state, (b) shows more specific
phenomenon than (a) due to the coherent quantum
superposition of wave and particle state.}
\end{figure}

We think it is important to study the intermediate behavior
between wave and particle for revealing the nature of photon essentially. For this purpose, it firstly should be
required to prepare the pure superposition state of wave and
particle. Now, we insert the device in rounded rectangle of Fig.~2.
to research the genuine superposition behavior of wave and
particle state. After the photon passes through the HWPs, the
photon state will be involved as
\begin{eqnarray}\label{e5}
|\psi''\rangle=\frac{1}{\sqrt{2}}[(\cos\alpha|\mathrm{particle}\rangle+\sin\alpha|\mathrm{wave}\rangle)|H\rangle
-(\cos\alpha|\mathrm{particle}\rangle-\sin\alpha|\mathrm{wave}\rangle)|V\rangle].
\end{eqnarray}
Hence $\mathrm{D_{1}}$ and $\mathrm{D_{4}}$ ($\mathrm{D_{2}}$ and
$\mathrm{D_{3}}$) will detect wave-particle superposition state
$\cos\alpha|\mathrm{particle}\rangle-\sin\alpha|\mathrm{wave}\rangle$
($\cos\alpha|\mathrm{particle}\rangle+\sin\alpha|\mathrm{wave}\rangle$)(not
normalized). And we can obtain pure wave-particle superposition by
postselection. Take $\mathrm{D_{2}}$ and $\mathrm{D_{3}}$ for
example, the normalized wave-particle superposition state obtained
in these outputs is accurately expressed as
\begin{eqnarray}\label{e6}
|\phi\rangle=\frac{1}{\sqrt{1+\sqrt{2}\sin\alpha\cos\alpha\cos\theta}}(\cos\alpha|\mathrm{particle}\rangle+\sin\alpha|\mathrm{wave}\rangle).
\end{eqnarray}
For this state, we can theoretically derive the coincidence count
probability (or interference pattern) of path 0 (detector
$\mathrm{D_{2}}$),
\begin{eqnarray}\label{e7}
I'_{0}(\theta,\alpha)=\frac{1+\sin^{2}\alpha\cos\theta+\sqrt{2}\sin2\alpha\cos^{2}\frac{\theta}{2}}{2+\sqrt{2}\sin2\alpha\cos\theta},
\end{eqnarray}
which is very different from the probability of mixed states in
Eq.~(4). In order to distinctly compare them, we plot the
probability distribution $I'_{0}(\theta,\alpha)$ as shown in
Fig.~3(b). Comparing with Fig.~3(a), Fig.~3(b) is more
unimaginable, which should be the essential behavior of a photon
and shows the genuine morphing behavior between wave and particle.
The only overlap between the two cases is that when $\alpha=0$ and
$\frac{\pi}{2}$, both of the two graphs correspond to
particle-like and wave-like behaviors, respectively. In
particular, it is worth to note that when $\alpha=\frac{\pi}{4}$,
the photon is in the superposition state
$|\psi\rangle=\frac{1}{\sqrt{2+\sqrt{2}\cos\theta}}(|\mathrm{particle}\rangle+|\mathrm{wave}\rangle)$,
but the probability distribution is a horizontal straight line and
doesn't change with phase shift $\theta$ in Fig.~3(b).
Theoretically, when $\alpha$ takes values $(2n+\frac{1}{4})\pi$
and $(2n+\frac{3}{4})\pi$, $(n=0,1,2...)$, $I'_{0}(\theta)$ will
equal to $\frac{1}{4}(2+\sqrt{2})$ and $\frac{1}{4}(2-\sqrt{2})$,
respectively. In this case, the visibility $\mathcal{V}=0$, and
the photon is in the wave-particle superposition state but only
shows the particle-like behavior, which is the
quantum interference effect resulting from the coherence between wave and
particle.

\section{Discussion and conclusion}

So far, we have analyzed the quantum delayed-choice experiment in
detail and pointed out that experimenter can observe which
behavior photon exhibits as soon as it is detected by a detector,
while the photon still doesn't know if it is supposed to be a wave
or a particle. Based on the basic idea of quantum delayed-choice experiment, we have proposed a simper linear optical scheme for realizing
quantum delayed-choice experiment, which only requires the most
common optical elements in optics laboratory\cite{17,18}, such as
BS, PBS, phase shifter, and HWP. Hence the present scheme can be
easily realized under the current experimental condition. The
scheme allowed us to selectively observe the wave-particle mixture
or superposition behavior by choosing different detection device.
In addition, quantum delayed-choice experiment allows wave behavior and particle behavior to be observed with a single experimental setup by postselection, and most of previous works has studied the wave-particle mixture behavior. In this paper, we can see wave-particle mixture behavior is very different from the superposition behavior, which may be more closed to the photon's essential behavior and needed to be test experimentally in the future. 

In conclusion, we have proposed a simple realizable implementation scheme of
quantum delayed-choice experiment without initial entangled resource, which can be used for
selectively observing different behaviors of photons. And the genuine wave-particle
superposition behavior has been shown for the first time in this
paper. Especially, we have explicitly compared the wave-particle
mixture behavior with the wave-particle superposition behavior and
found the quantum interference effect between wave and particle behavior, which may be meaningful to
reveal the nature of photon essentially.

\begin{center}$\mathbf{Acknowledgments}$\end{center}

This work is supported by the National Natural Science Foundation
of China under Grant Nos. 61068001 and 11264042; China
Postdoctoral Science Foundation under Grant No. 2012M520612; the
Program for Chun Miao Excellent Talents of Jilin Provincial
Department of Education under Grant No. 201316; and the Talent
Program of Yanbian University of China under Grant No. 950010001.

\end{document}